\documentclass[]{aa}

\usepackage{graphicx}

\newcommand{\ie}{\mbox{i.e.}}%
\newcommand{\eg}{\mbox{e.g.}}%
\newcommand{\etal}{\mbox{et al.\ }}%
\newcommand{\tpm}{\ensuremath{\pm}}%
\newcommand{\tsim}{\ensuremath{\sim}}%
\newcommand{\tsimeq}{\ensuremath{\simeq}}%
\newcommand{\lam}{\ensuremath{\lambda}}%
\newcommand{\Ha}{\ensuremath{\mbox{H}\alpha}}%
\newcommand{\Hb}{\ensuremath{\mbox{H}\beta}}%
\newcommand{\Hg}{\ensuremath{\mbox{H}\gamma}}%
\newcommand{\apj}{ApJ\ }%
\newcommand{\aap}{A\&A\ }%
\newcommand{\iauc}{IAU Circ.\ }%

\begin{document}

\thesaurus{6
        ( 08.19.4;		
          08.19.5 SN1987A;	
          09.19.2;		
          02.18.7		
        )}

\msnr{H2633}

\title{The late-time expansion of the ejecta of SN~1987A\footnotemark[1]}

\author{Rolf A. Jansen \and Peter Jakobsen}
\offprints{R.\ A. Jansen ({rjansen@}\-{astro.}\-{estec.}\-{esa.nl})}

\institute{Astrophysics Division, Space Science Department of ESA,
           ESTEC, NL-2200 AG Noordwijk, The Netherlands\\
           email: rjansen or pjakobse@astro.estec.esa.nl
}

\date{Received 9 January 2001 / Accepted 25 February 2001}

\authorrunning{R.\ A. Jansen \& P. Jakobsen}
\titlerunning{Evolution of the ejecta of SN~1987A}

\maketitle

\renewcommand{\thefootnote}{$\fnsymbol{footnote}$}
\footnotetext[1]{Based on observations made with the NASA/ESA
                \emph{Hubble Space Telescope}, obtained at the Space
                Telescope Science Institute (STScI) and from the data
                archive at the STScI, which is operated by AURA, Inc.,
                under NASA contract NAS 5-26555}
\renewcommand{\thefootnote}{$\arabic{footnote}$}

\markboth{R.\ A. Jansen \& P. Jakobsen}
         {Expansion of the ejecta of SN~1987A}


\begin{abstract}

The evolution of the shape and size of the ejecta of SN~1987A is
analyzed over a period of \tsim8 years based on \emph{HST} images and
spectra taken between 1278 and 4336 days after the supernova outburst. 
We combine both proprietary and archival \emph{HST} data obtained with
the FOC, WFPC2 and STIS.  The low resolution near-UV prism FOC spectrum
obtained at day 3043 has not been described previously.  Although the
FWHM of the ejecta grew linearly over the time span studied, the
appearance of the SN envelope also changed markedly with wavelength.  At
visible wavelengths (\lam \tsimeq 5000~\AA) the ejecta became
progressively more elongated, reaching an ellipticity of $\epsilon\simeq
0.25$ by day 4000.  In the near-UV (\lam \tsimeq 2500~\AA), the ejecta
remained closely circular ($\epsilon\leq 0.1$) and \tsim50\% larger in
angular extent than in the visible.  The FOC prism observations show
that the large extent of the SN envelope is confined to a grouping of
resonance lines spanning \ion{Mg}{i}~\lam2852,
\ion{Mg}{ii}~\lam\lam2795,2802\AA\ and several \ion{Fe}{ii} multiplets
-- thereby confirming that the larger size of the debris in the near-UV
is due to scattering in these optically thick transitions compared to
the optically thin forbidden and semi-forbidden transitions that
dominate the visible spectrum.  The available data are not of sufficient
quality to detect the slight deviation from linear expansion expected
for the outermost regions of the near-UV images as predicted by Chugai
\etal (1997). 

\keywords{supernovae: individual (SN~1987A) --- supernova remnants}

\end{abstract}

  
\section{Introduction}

The remnant of SN~1987A and its surroundings have been extensively
monitored with the \emph{Hubble Space Telescope} (\emph{HST}) since its
commissioning, first with the aberated telescope, later with WFPC2 and
the COSTAR-corrected FOC, and most recently using STIS.  While much
attention has been devoted to the study of the inner circumstellar ring
(\eg, Panagia \etal 1991; Luo, McCray, \& Slavin 1994; Plait \etal 1995;
Crotts \etal 1995) and its interaction with the SN blast wave (\eg,
Sonneborn \etal 1998; Crotts \& Heathcote 2000; Michael \etal 2000), the
evolution of the spatially resolved SN ejecta itself is also of
considerable interest in its own right. 

Jakobsen \etal (1991) demonstrated that the expanding envelope of
SN~1987A was spatially resolved by \emph{HST} already in the first FOC
images taken 1278 days after the explosion.  These data combined with
subsequent FOC observations made on days 1754, 2511 and 2533 showed that
the outer envelope of SN~1987A was expanding linearly with time in a
self-similar fashion (Jakobsen \etal 1994). 

The early FOC observations also revealed that the size of the SN debris
varies significantly with wavelength, appearing twice as large in the
near-UV (F275W filter) compared to the visible (F501N [\ion{O}{iii}]
filter).  Jakobsen \etal (1993) suggested that this difference in
apparent size is likely due to an opacity effect, with optical depth
unity in the near-UV being reached further out in the expanding ejecta
compared to the optically thin [\ion{O}{iii}] line which probes deeper
into the expanding envelope.  This interpretation was further refined
based upon FOS observations obtained by Wang \etal (1996) and Chugai
\etal (1997), who showed that the near-UV (\lam\lam2350-2900) spectrum
of SN~1987A spanned by the FOC F275W filter is dominated by a dense
grouping of resonance lines containing \ion{Mg}{i}~\lam2852 and
\ion{Mg}{ii}~\lam\lam2795,2802 and several multiplets of \ion{Fe}{ii}. 
These authors also showed that the velocity-widths of the
\ion{Mg}{i}~\lam2852 and \ion{Mg}{ii}~\lam\lam2795,2802 lines are in
good agreement with the expansion velocities inferred from the FOC
near-UV images, and considerably broader than those of the forbidden and
semi-forbidden lines which dominate the visible spectrum. 

In this paper we present previously unpublished FOC objective prism
observations of SN~1987A taken on day 3043 which, by bringing together
both imaging and spectroscopic information serve to further explore the
cause of the wavelength dependence of the apparant size of SN~1987A. 
Our main finding is that the largest extent of the ejecta is indeed
confined to the above grouping of high opacity near-UV resonance lines,
thereby confirming this explanation for the change in apparent size with
wavelength.  We also compare these new FOC angular diameter measurements
to matching measurements derived from archival WFPC-2 and STIS data
spanning up to day 4336.  We show that, provided care is taken to put
the different measures onto a common system by taking the non-zero
ellipticity of the SN~1987A envelope into account, the data show good
agreement and are consistent with linear expansion of the SN~1987A
envelope over the entire \tsim8 year time span probed by the publicly
available \emph{HST} observations. 

At the time of writing, some 14 years after the explosion, the afterglow
of SN~1987A proper has faded beyond the reach of \emph{HST} -- and is
likely to remain so even when the Advanced Camera for Surveys is
installed.  This paper is therefore an attempt at summarizing the
available \emph{HST} data on the shape and expansion of the SN~1987
debris.


\section{Data}

\subsection{Day 3043 FOC Observations}

Low resolution spectra of the remnant of SN~1987A were obtained using
the COSTAR-corrected FOC on board \emph{HST} on 1995 June 24
(JD~2,449,892; day 3043).  The FOC was used in filterless mode with the
F/96 near-UV objective prism inserted.  A total exposure of 16683~s was
obtained, divided over 12 exposures of 1221.5--1432.5~s each, with
pointings dithered in a $0\farcs 2$ pattern for optimal removal of
reseau marks and flatfield features not taken out by the standard
``pipeline'' procedures. 

With the $512\times1024$ pixel detector format used, non-linearity in
extended objects sets in at a count rate of 0.08
counts~sec$^{-1}$~pix$^{-1}$ and saturation at 0.37
counts~sec$^{-1}$~pix$^{-1}$.  The count rates in the spectrum of the SN
remnant are within the linear regime even at \Ha\ and [\ion{O}{iii}],
but the optical portions of companion stars 2 and 3 (in the enumeration
of West \etal 1987) are severely saturated due to the low dispersion of
the objective prism at the longer wavelengths.  The \emph{HST}
roll-angle (V3 position angle) during these observations was 15.2
degrees, causing the spectrum of companion star 2 to overlap the
circumstellar ring at the position of the faint foreground star (Plait,
Chevalier \& Kirshner 1992).

\begin{figure}[t!]
\vspace*{4pt}
\centerline{
   \hfill\includegraphics[width=0.485\textwidth]{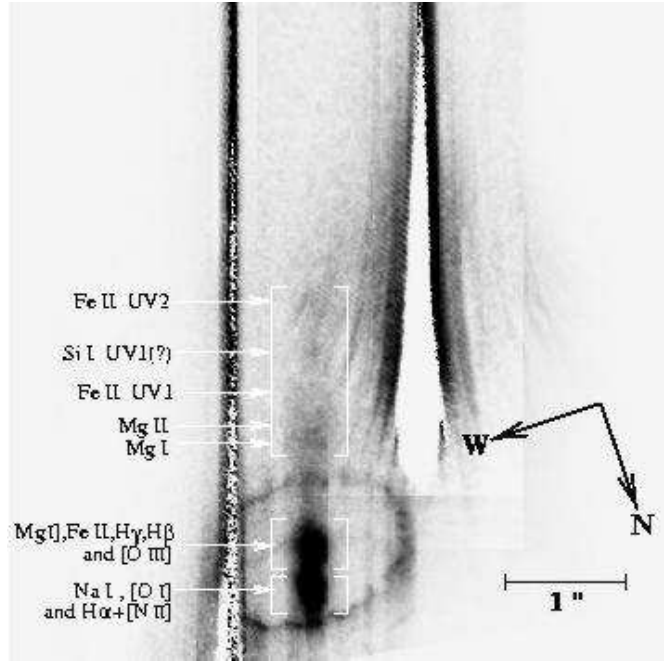}\hfill
}\par
\parbox[t]{0.485\textwidth}{
   \caption[]{
Greyscale rendition of the portion of the FOC objective prism image that
includes the dispersed image of the remnant of SN~1987A and the 
circumstellar ring in [\ion{O}{iii}] and (much fainter) in 
\Ha+[\ion{N}{ii}].  The spatial image scale and the image orientation
are indicated.  The spectra from companion stars 2 and 3 (located 
$2\farcs9$ NW and $1\farcs6$ SE of the SN remnant, respectively) have 
been partially removed.  The \lam\lam2350--2900\AA\ portion of the SN
spectrum spanned by the grouping of \ion{Mg}{i}/\ion{Mg}{ii}/\ion{Fe}{ii}
near-UV resonance lines (see text) is indicated.
   }
}
\end{figure}
%

The individual exposures were reduced using the standard \emph{HST}
postprocessing software and subsequently combined, weighted by exposure
time.  To reduce the contamination of the SN remnant spectrum by the
near-UV light of star 2, located $2\farcs9$ NW of SN 1987A, we used a
matched (in wavelength) portion of the spectrum of star 3, located
$1\farcs6$ to the SE, as a PSF template.  None of the stars in the field
of view for which we observe the optical portion of the spectrum
aproximates the brightness of star 3.  For subtraction of star 3 we
therefore had to use a fainter star, located $7\farcs 9$~S of SN~1987A
as a template.  Although the wings of the optical PSF in the template
are not sampled to sufficiently faint levels to provide a good match,
subtraction of this scaled-up template does significantly reduce the
contamination from star 3 at the position of the SN remnant spectrum. 
In the following analysis, residual light from star 3 is the dominant
source of error in the SN remnant spectrum for wavelengths shortward of
3000\AA.

\begin{figure*}[t!]
\centerline{
   \includegraphics[width=\textwidth]{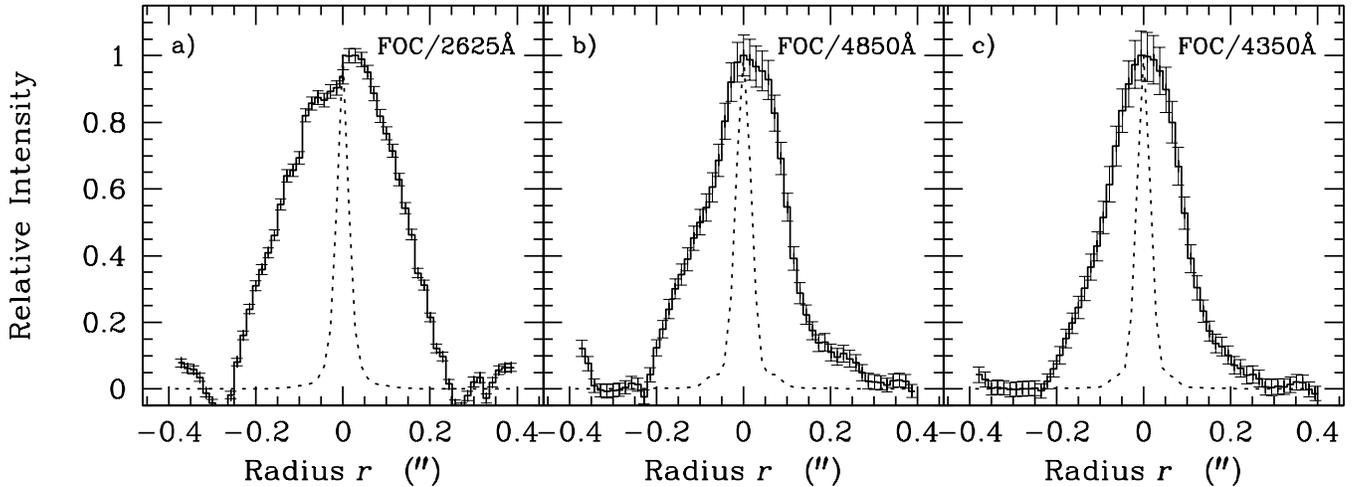}
}
\caption[]{
Radial intensity profiles of the ejecta of SN~1987A in the day 3043 FOC
data for (\emph{a}) the near-UV measured in a 550\AA\ wavelength
interval centered on 2625\AA\ (comparable to the earlier F275W data and
to more recent F255W filter WFPC2 data), and (\emph{b}) the optical in a
1300\AA\ interval centered on 4850\AA\ (comparable to the earlier F501N
data [see text] and more recent F555W filter WFPC2 data), and (\emph{c})
a 700\AA\ interval centered on 4350\AA\ (matched to the WFPC2 F439W
filter). The intensity profiles were normalized to the peak intensity. 
In each figure, the FOC PSF for the central wavelength of the interval
is overlayed for comparison ({\it dotted}\/). 
}
\end{figure*}
%

The portion of the co-added and partially subtracted FOC image showing
the (slitless) near-UV prism spectrum of SN~1987A is shown in Figure~1. 
In this exposure, the circumstellar ring appears bright in the light of
[\ion{O}{iii}] and is faintly discernable in \Ha+[\ion{N}{ii}].  The
optical portion of the SN spectrum shows clear features, roughly located
at [\ion{O}{iii}] and \Ha+[\ion{N}{ii}], but resulting from blends of
the lines of \ion{Mg}{i}], \ion{Fe}{ii}, \Hg, \Hb, and [\ion{O}{iii}] in
the former and \ion{Na}{i}, [\ion{O}{i}] and \Ha+[\ion{N}{ii}] in the
latter case.  The SN spectrum is traceable down to a wavelength of
\tsim\lam2200~\AA\ in the near-UV.  Moreover, consistent with the
previous FOC imaging observations, the near-UV spectrum is noticeably
more spatially extended than the visible spectrum. 

In order to explore this quantitatively we selected two pixel regions
centered on the SN remnant spectrum in the spatial direction and on
wavelengths 2625\AA\ and 4850\AA\ in the dispersion direction.  Along
the dispersion direction we averaged the pixel values in 93 lines and 26
lines, corresponding to 550\AA\ and 1300\AA\ wavelength intervals,
respectively.  This choice was a compromise between encompassing the
respective \ion{Mg}{ii}/\ion{Mg}{i}/\ion{Fe}{ii} and
\ion{Mg}{i}]/\Hb/\ion{Fe}{ii}/[\ion{O}{iii}] features (see below), and
gathering enough signal to reduce the noise while avoiding contamination
from the underlying image of the ring.  The resulting radial intensity
profiles are presented in Figs.~2\emph{a} and 2\emph{b}.  The FOC PSF at
2625\AA\ and 4850\AA\ is overlayed in each figure.  The apparent size of
the SN debris as measured by the FWHM of the radial intensity profiles
is 301\tpm7~mas at 2625\AA\ and 201\tpm10~mas at 4850\AA.  These values
are corrected for the PSF.  We also selected a third, 20 pixel (700\AA)
wide region centered at 4350\AA, chosen to match a grouping of
\ion{Fe}{ii} lines and the semi-forbidden \ion{Mg}{i}] line (see
Sect.~2.2.1).  By sampling the same species (Mg and Fe), a comparison
between the NUV and 4350\AA\ line groupings is likely to be cleaner than
a comparison with the 4850\AA\ band, or even with the [\ion{O}{iii}]
line.  The apparent size (FWHM) of the debris measured in the radial
intensity profile at 4350\AA\ (Fig.~2\emph{c}) is 188\tpm13~mas,
smallest of any wavelength sampled in these data.

\subsection{Archival HST data}

The data on SN~1987A contained in the public \emph{HST} archive consists
of a large number of multifilter WFPC2 images, as well as STIS and FOS
near-UV and optical spectra, and the earlier FOC imaging observations. 
The available data are summarized in Table~1.  References to published
observations are given in the final column of the table.  A large
portion of these data were obtained as part of the Supernova INtensive 
Study (SINS) project.  Of the WFPC2 data, we only use the Planetary
Camera chip (in the following referred to as PC2) which contains the
image of the SN remnant itself.

\begin{table*}[t!]
\footnotesize
\setlength{\tabcolsep}{3pt}
\begin{center}
\parbox[b]{\linewidth}{
   \caption[]{Overview of \emph{HST} Data on SN~1987A used in this paper.}
}
\begin{tabular*}{\linewidth}[h!]{ l l l l l r l l l }
\hline\noalign{\smallskip} 
\multicolumn{1}{c}{Day$^\dag$}      & \multicolumn{1}{c}{Date}             &
\multicolumn{1}{c}{Instrument}      & \multicolumn{1}{l}{Data type}        &
\multicolumn{1}{l}{Filter}          & \multicolumn{1}{c}{Exposure$^\ddag$} &
\multicolumn{1}{c}{Proposal$^\ast$} & \multicolumn{1}{c}{P.I.}             &
\multicolumn{1}{c}{Reference}       \\
\noalign{\smallskip}
\hline\noalign{\smallskip}
1278 & Aug 23 1990 & FOC   & Imaging & F275W     &  838 & 2999 & F. Macchetto & Jakobsen et al. 1991\\[-1pt]
     &             &       &         & F501N     & 1660 &      & & \\[-1pt]
1754 & Dec 13 1991 & FOC   & Imaging & F275W     & 2200 & 3874 & F. Macchetto & Jakobsen et al. 1993\\[-1pt]
     &             &       &         & F501N     & 2200 &      & & \\[-1pt]
2511 & Jan 08 1994 & FOC   & Imaging & F275W+ND1 & 4784 & 5186 & W. Sparks    & Jakobsen et al. 1994\\[-1pt]
     &             &       &         & F501N     & 4784 &      & & \\[-1pt]
2533 & Jan 30 1994 & FOC   & Imaging & F275W     & 3586 & 5152 & P. Jakobsen  & Jakobsen et al. 1994\\[-1pt]
     &             &       &         & F501N     & 3586 &      & & \\[-1pt]
2537 & Feb 03 1994 & PC2   & Imaging & F255W     & 2400 & 5203 & J. Trauger   & Burrows et al. 1995\\[-1pt]
     &             &       &         & F502N     & 2400 &      & & \\[-1pt]
     &             &       &         & F547M     & 2400 &      & & \\[-1pt]
     &             &       &         & F656N     & 2400 &      & & \\[-1pt]
2770 & Sep 24 1994 & PC2   & Imaging & F255W     & 1800 & 5753 & R. Kirshner  & Panagia et al. 1996\\[-1pt]
     &             &       &         & F336W     & 1200 &      & & \\[-1pt]
     &             &       &         & F439W     &  800 &      & & \\[-1pt]
     &             &       &         & F502N     & 4800 &      & & \\[-1pt]
     &             &       &         & F555W     &  600 &      & & \\[-1pt]
     &             &       &         & F658N     & 4800 &      & & \\[-1pt]
     &             &       &         & F675W     &  600 &      & & \\[-1pt]
     &             &       &         & F814W     &  600 &      & & \\[-1pt]
2875 & Jan 07 1995 & FOS & Spectroscopy & G190H  & 5610 & 5753 & R. Kirshner & Chugai et al. 1997\\[-1pt]
     &             &       &         & G270H     & 6000 &      & & \\[-1pt]
     &             &       &         & G400H     & 4000 &      & & \\[-1pt]
     &             &       &         & G570H     & 4000 &      & & \\[-1pt]
2932 & Mar 05 1995 & PC2   & Imaging & F439W     &  800 & 5753 & R. Kirshner  & \\[-1pt]
     &             &       &         & F555W     &  600 &      & & \\[-1pt]
     &             &       &         & F675W     &  600 &      & & \\[-1pt]
     &             &       &         & F814W     &  600 &      & & \\[-1pt]
3043 & Jun 24 1995 & FOC & Spectroscopy & NUV-prism & 16200 & 6130 & P. Jakobsen & (this paper) \\[-1pt]
3270 & Feb 06 1996 & PC2   & Imaging & F255W     & 2300 & 6020 & R. Kirshner  & \\[-1pt]
     &             &       &         & F336W     & 1200 &      & & \\[-1pt]
     &             &       &         & F439W     &  950 &      & & \\[-1pt]
     &             &       &         & F502N     & 7800 &      & & \\[-1pt]
     &             &       &         & F555W     &  600 &      & & \\[-1pt]
     &             &       &         & F658N     & 5200 &      & & \\[-1pt]
     &             &       &         & F675W     &  600 &      & & \\[-1pt]
     &             &       &         & F814W     &  600 &      & & \\[-1pt]
3693 & Apr 04 1997 & STIS  & Imaging & OII       & 2415 & 7122 & G. Sonneborn & \\[-1pt]
3790 & Jul 10 1997 & PC2   & Imaging & F255W     & 2600 & 6437 & R. Kirshner  & Garnavich et al. 1997a\\[-1pt]
     &             &       &         & F336W     & 1600 &      & & \\[-1pt]
     &             &       &         & F439W     &  800 &      & & \\[-1pt]
     &             &       &         & F555W     &  600 &      & & \\[-1pt]
     &             &       &         & F656N     & 5600 &      & & \\[-1pt]
     &             &       &         & F675W     &  600 &      & & \\[-1pt]
     &             &       &         & F814W     &  800 &      & & \\[-1pt]
3792 & Jul 12 1997 &       &         & F502N     & 8200 &      & & \\[-1pt]
3878 & Oct 06 1997 & STIS  & Spectroscopy & G430L &5095 & 7434 & R. Kirshner  & Garnavich et al. 1997b\\[-1pt]
3879 & Oct 07 1997 &       &         & G750M     & 4772 &      & & \\[-1pt]
3880 & Oct 08 1997 &       &         & G230L     &14082 &      & & \\[-1pt]
4336 & Jan 07 1999 & PC2   & Imaging & F255W     & 7200 & 7434 & R. Kirshner  & Garnavich et al. 1999\\[-1pt]
     &             &       &         & F336W     & 2400 &      & & \\[-1pt]
     &             &       &         & F439W     & 1000 &      & & \\[-1pt]
     &             &       &         & F555W     &  600 &      & & \\[-1pt]
     &             &       &         & F656N     & 7200 &      & & \\[-1pt]
     &             &       &         & F675W     & 1200 &      & & \\[-1pt]
     &             &       &         & F814W     &  800 &      & & \\[-1pt]
\noalign{\smallskip}\hline\noalign{\smallskip}
\end{tabular*}
\parbox[b]{\linewidth}{
   \noindent {\sc Notes:}
   $^\dag$Since the SN outburst on Feb 23, 1987;
   $^\ddag$Total exposure time in seconds;
   $^\ast$ID number of the observing program.
}
\end{center}
\vspace*{-5mm}
\end{table*}
%

\subsubsection{Spectroscopic observations and line identification}

In Fig.~3 we reproduce the 1995 January~7 (Day~2875) FOS spectrum
presented in Chugai \etal (1997).  The bandpasses of selected \emph{HST}
filters are overlayed to show which spectral features contribute most to
the total signal in imaging observations through these filters.  The
identifications of the emission features were taken from Wang \etal
(1996) and Chugai \etal (1997). 

In the day 3880 STIS observations (1997, October 8) the 2250--3000\AA\
wavelength region is very similar in appearance to the corresponding
portion of the 1995 January 7 FOS spectrum of Fig.~3.  The NUV features
seen in these higher resolution spectra can also be gleaned in the
dispersed 1995 June 24 FOC near-UV prism image (Fig.~1).

\begin{figure*}[ht!]
\centering
\includegraphics[width=0.68\textwidth]{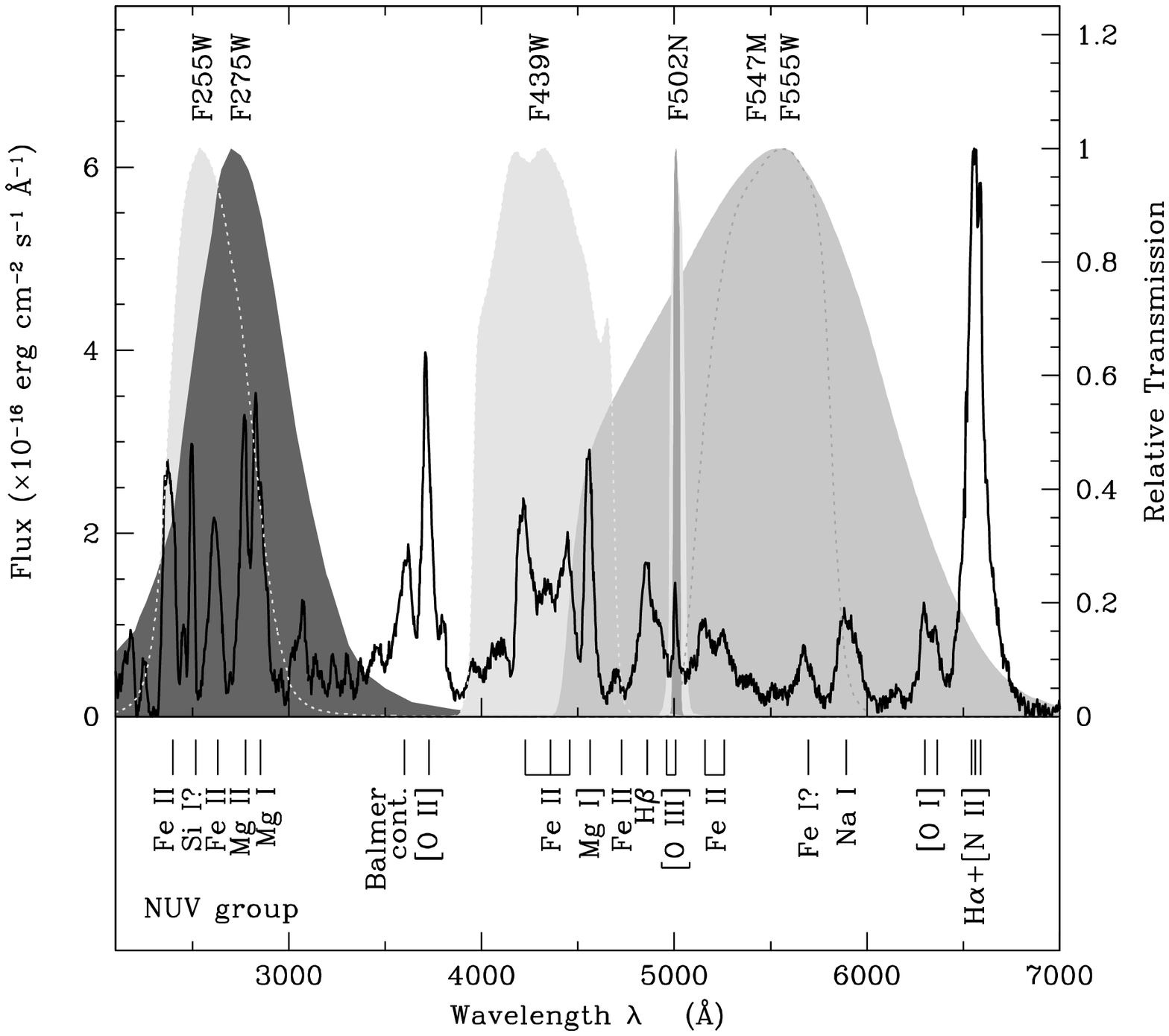}
\caption[]{
Match of selected NUV and visible \emph{HST} filters to spectral
features in the SN~1987A spectrum.  The spectrum shown is the day 2875
FOS spectrum presented by Chugai \etal (1997).  The line identifications
are from that paper and from Wang \etal (1996). The transmission axis
(right) refers to the filters shown. }
\end{figure*}
%

The 2350--2900\AA\ interval used to extract the radial intensity profile
of Fig~2\emph{a} selects the strong NUV resonance lines of \ion{Mg}{ii},
\ion{Mg}{i} and \ion{Fe}{ii}.  The F255W (PC2) and F275W (FOC) filters
are a good match to this ensemble of lines, although the latter places
slightly more weight on the Mg lines.  The narrow band filters F501N
(FOC) and F502N (PC2) sample emission from [\ion{O}{iii}] and
\ion{Fe}{ii} \lam5018\AA.  The F555W (PC2) filter samples a multitude of
lines: \ion{Mg}{i}], \Hb, [\ion{O}{iii}], \ion{Fe}{ii}, \ion{Na}{i},
[\ion{O}{i}], and \Ha+[\ion{N}{ii}] in its red tail.  Nonetheless,
comparisons of F501N and F502N filter data with data in this filter are
still meaningful, since the dominant lines spanned by the F555W filter
are all low opacity forbidden and semi-forbidden transitions. 

In the following we will treat the FOC F275W and PC2 F255W data as
equivalent and sampling the resonance line dominated near-UV emission of
SN~1987A.  The FOC F501N and PC2 F502N, F555W and F547M imaging data are
grouped in a similar manner as sampling the forbidden and semi-forbidden
visible line emission from SN~1987A.


\section{Results}

\subsection{Ellipticity of the SN debris}

The FOC and PC2 imagery reveal that the shape of the SN debris in the
visible became noticeably elongated after about day 2500.  The later day
3792 F502N and the day 4336 F555W data show apparent axis ratios close
to 0.7, whereas the earliest day 1278 and 1754 FOC data were consistent
with negligible elongation (Jakobsen \etal 1993).  We limit ourselves in
the present discussion to the large scale structure of the SN debris. 
To first order the shape of the debris can be described as elliptical. 
It is obvious from the best S/N images, especially in the narrow-band
filters, that there is fine-structure in the debris.  Pun \& Kirshner
(1996) tentatively described the debris shape as a dumbell-shaped
structure.  Accurate description of these features and monitoring of
their evolution over time, however, would have required excessively long
exposure times to obtain the necessary signal.

\begin{figure*}[t!]
\centerline{
   \includegraphics[width=0.40\textwidth]{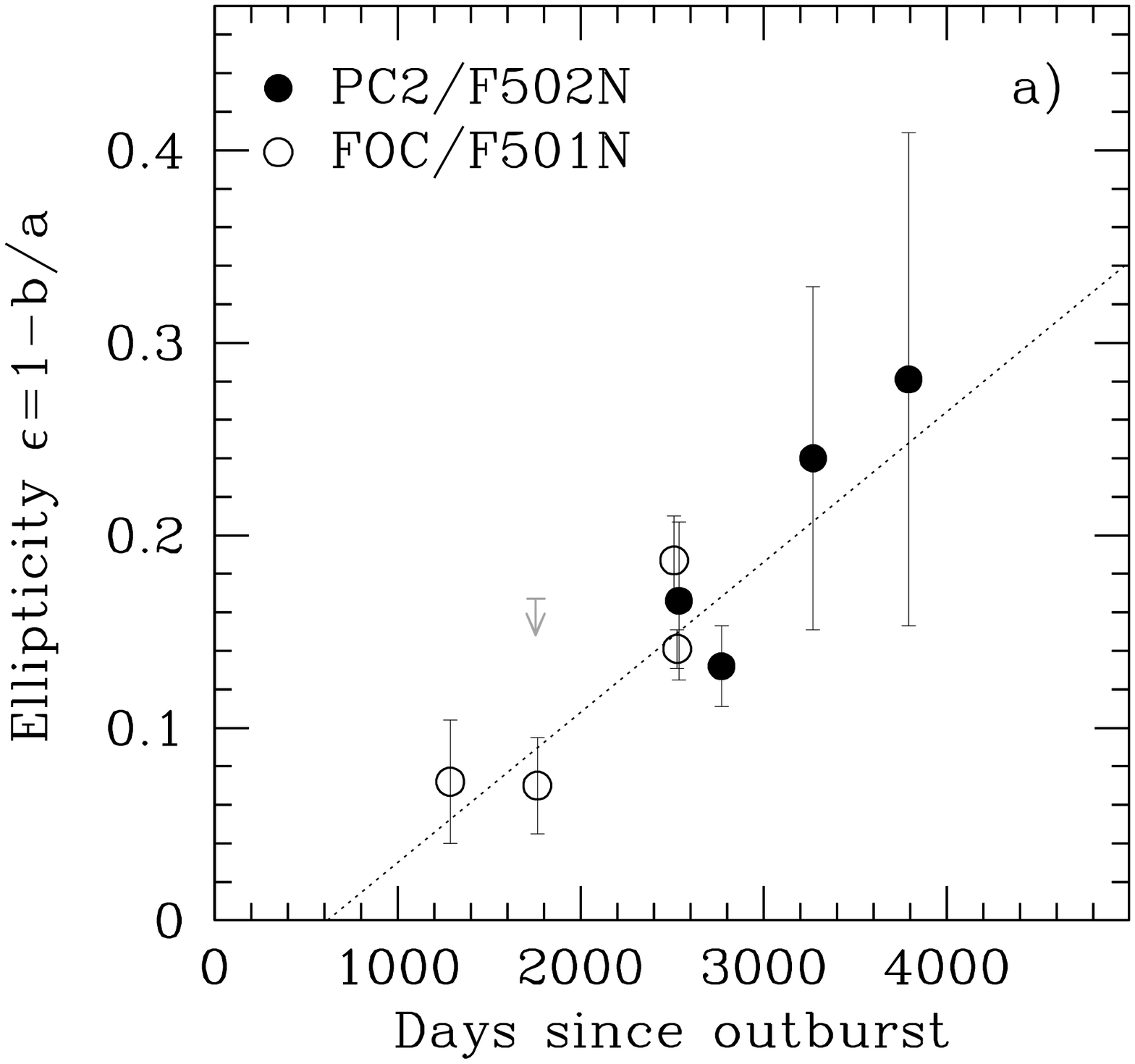}$\quad $
   \includegraphics[width=0.40\textwidth]{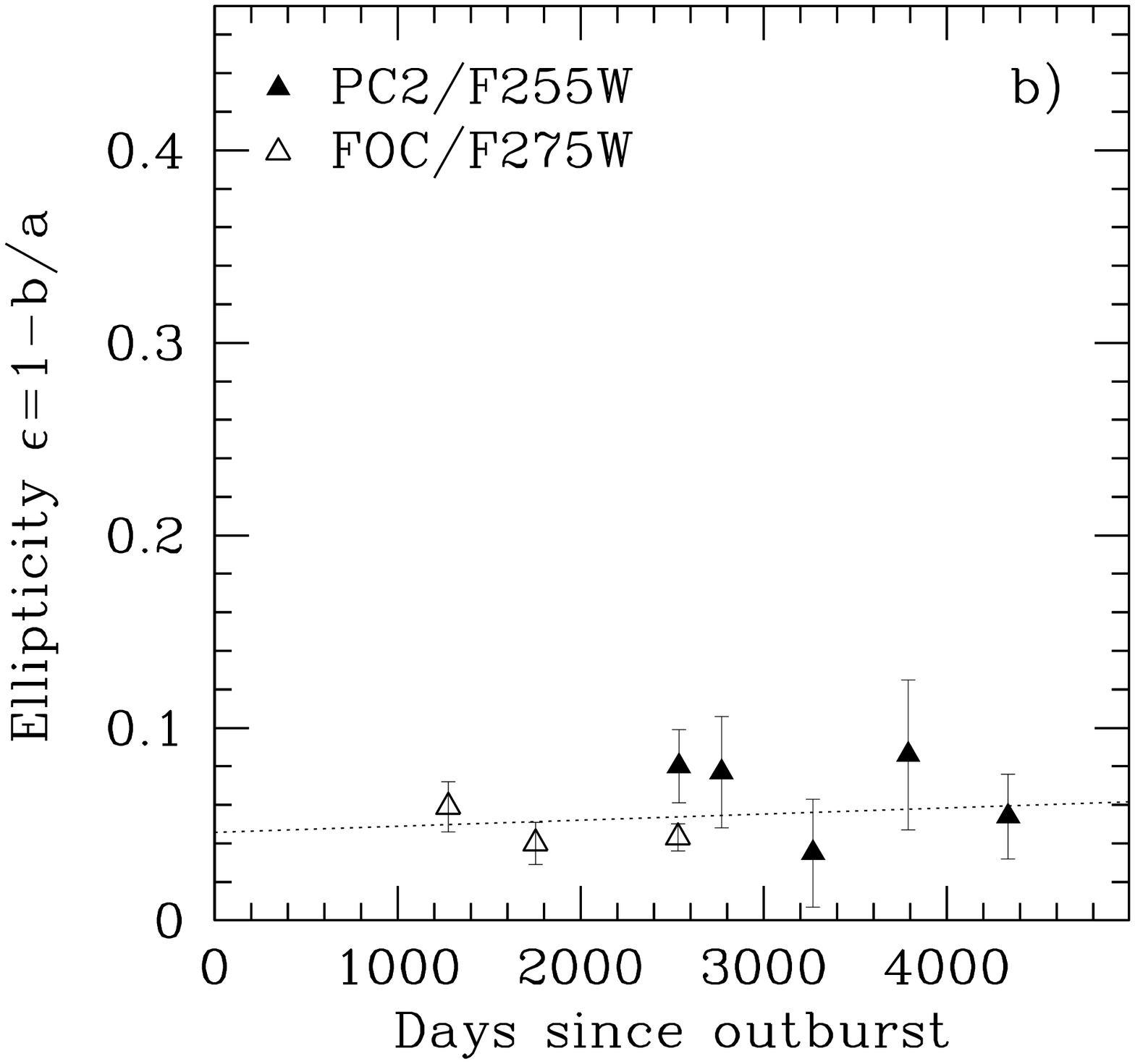}
}
\caption[]{
Time evolution of the ellipticity $\epsilon=1-b/a$ of the debris of
SN~1987A as a function of wavelength.  \emph{a}) Measurements of
$\epsilon$ in the F501N (FOC) and F502N (PC2) filters with $\pm 1\sigma$
error bars.  The formal upper limit from Jakobsen \etal 1993 for the day
1754 F501N data is indicated by the grey arrow symbol.  A weighted
least-squares fit to the combined F501N and F502N data is overlayed. 
(\emph{b}) As (\emph{a}) for the F275W (FOC) and F255W (PC2) filters. 
}
\end{figure*}
%

We derived ellipticities from the 2nd order image moments measured in a
small image section centered on the SN remnant.  If we denote the 2nd
order image moments by $M_{xx}$, $M_{yy}$ and $M_{xy}$ then the
ellipticity is given by:\[ 
\epsilon = 1 - b/a = \frac{\sqrt{\left(M_{xx} - M_{yy}\right)^2 +
\left(2\,M_{xy}\right)^2}}{\left(M_{xx}+M_{yy}\right)}\quad .  \]
Initial measurements of the FWHM of the debris (see also Sect.~\ref{SNsize})
and the error thereon were used to define the minimum and maximum size
of the image region in which to measure the ellipticities.  The
ellipticity was measured in a set of nested image sections with
halfsizes that grow between FWHM$-$$\sigma_{\hbox{\scriptsize FWHM}}$
and FWHM$+$$\sigma_{\hbox{\scriptsize FWHM}}$ with increments of 1
pixel.  The adopted ellipticities used in the analysis below are the
weighted average of the individual measurements.  These tend to be
conservative in the sense that in the smaller image sections the
measured ellipticity tends to be systematically higher than in the
larger sections.

Fig.~4\emph{a} shows that the ellipticity in the [\ion{O}{iii}]
narrow-band filters (FOC/F501N and PC2/F502N) increases approximately
linearly with time.  A weighted linear least-squares fit gives a rate of
change $d\epsilon/dt$ of $(7.8\pm1.9)\times 10^{-5}$ day$^{-1}$ and
onset of the elongation at day $(600\pm400)$.  The large errors on the
PC2 F502N measurements are due in part to the larger plate scale of the
PC2 CCD (compared to the FOC camera), and in part to the low signal
levels in the SN debris, as these data were obtained primarily for the
purpose of studying the inner circumstellar ring. 

The debris shape is dramatically less elongated in the near-UV
(FOC/F275W and PC2/F255W) filters compared to the [\ion{O}{iii}] filter
(Fig.~4\emph{b}).  It is possible that in the near-UV the ellipticity
increases slowly with time, but low surface brightness, low-signal,
exposures preclude a definite measure.  A small constant ellipticity
$\epsilon = 1 - b/a \sim 0.05^{+0.03}_{-0.02}$ is consistent with the
available data. 

\subsection{Expansion of the SN Debris}
\label{SNsize}

In order to quantify the expansion of the envelope of the SN debris in a
consistent manner, we have to take its systematic change in elongation
at the later stages into account.  Also, considering the dimming and
hence lower S/N ratios of the more recent data, it is necessary to
characterize the size by as robust an estimator as possible.  We adopt
the FWHM of the intensity profile of the debris, but calculated in terms
of an elliptical or equivalent radius.  If the semi-major and semi-minor
axes at a given surface brightness level are denoted by $a$ and $b$,
respectively, then the elliptical radius $r$ corresponding to that
isophote is given by $r=\sqrt{ab}$.  Given an ellipticity
$\epsilon=1-b/a$ the elliptical radius is related to the semi-major axis
radius by $r=a\,\sqrt{1-\epsilon}$. 

For each epoch of observation and each filter, we performed aperture
photometry in a set of nested circular apertures and constructed radial
intensity profiles by differencing the signal in consecutive apertures. 
The FWHM measured in these radial intensity profiles was expressed in
terms of elliptical radii by multiplication with the geometric
correction term $\sqrt{1-\epsilon_{\lambda}(t)}$.  For
$\epsilon_{\lambda}(t)$ we adopted the empirical least-squares fits
shown for F501N/F502N and F255W/F275W filters in Figs.~4, and similar
fits for the other available filters.  Using a simple geometric
correction was possible because the spatial extent of the debris is much
larger than the PSF, and the distribution of the light is fairly smooth
and not very concentrated. 

In Figs.~5\emph{a} and \emph{b} we present the resulting measures of the
size of the SN debris as a function of time for both the visible (F501N,
F502N, F547M and F555W) and near-UV (F255W and F275W) data sets.  Also
plotted are the equivalent ``two dimensional'' spectroscopic measures
derived from the day 3043 FOC near-UV prism data shown in Fig.~2, and a
high-quality near-UV point derived in the same way from the day 3880
STIS G230L-grating spectrum.  In the case of the spectroscopic measures
we also applied a (small) geometric correction to account for the
mismatch of the position angles of the dispersion axis and the major
axis of the SN debris.

\begin{figure*}[ht!]
\centerline{
   \includegraphics[width=0.40\textwidth]{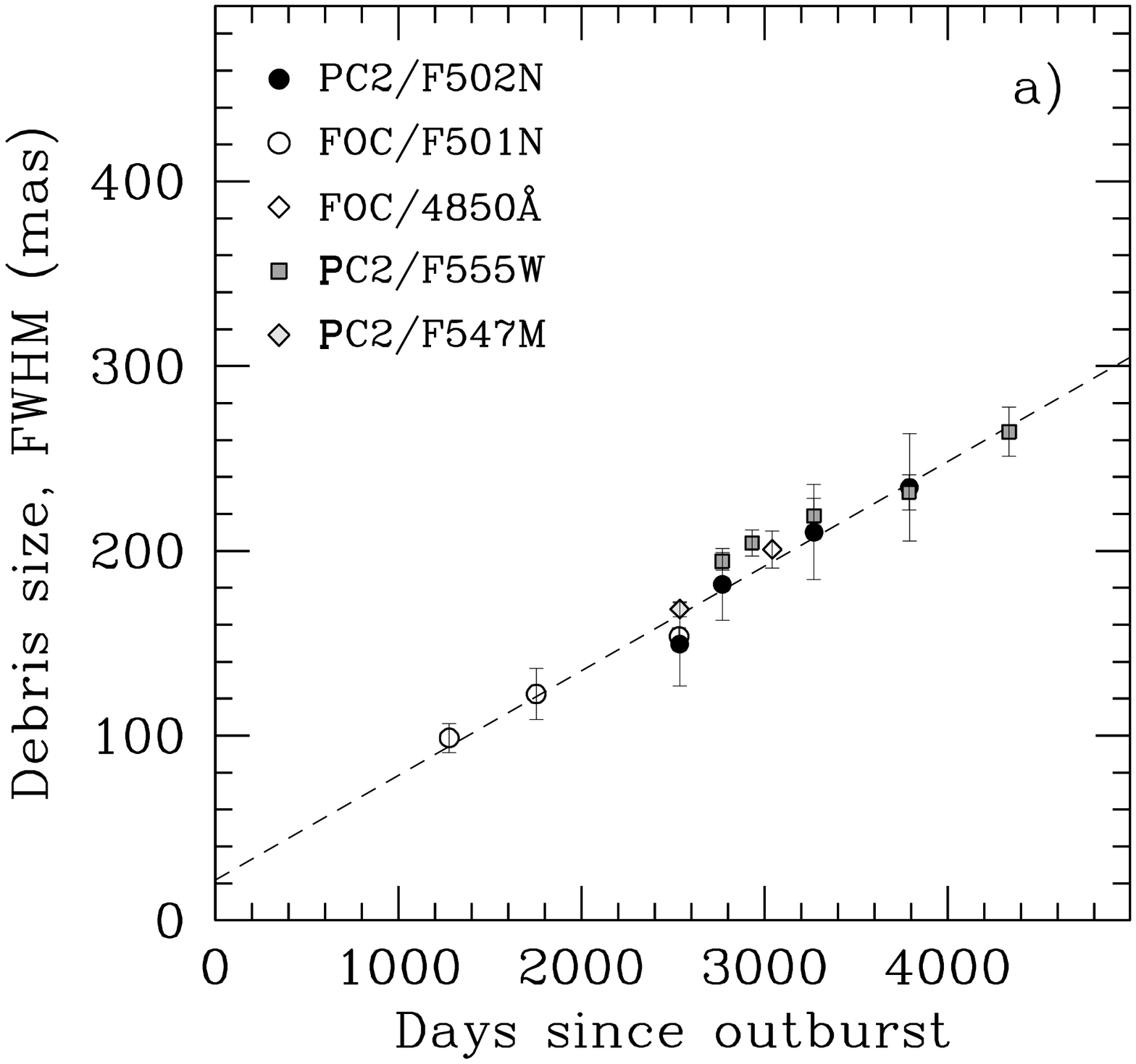}$\quad $
   \includegraphics[width=0.40\textwidth]{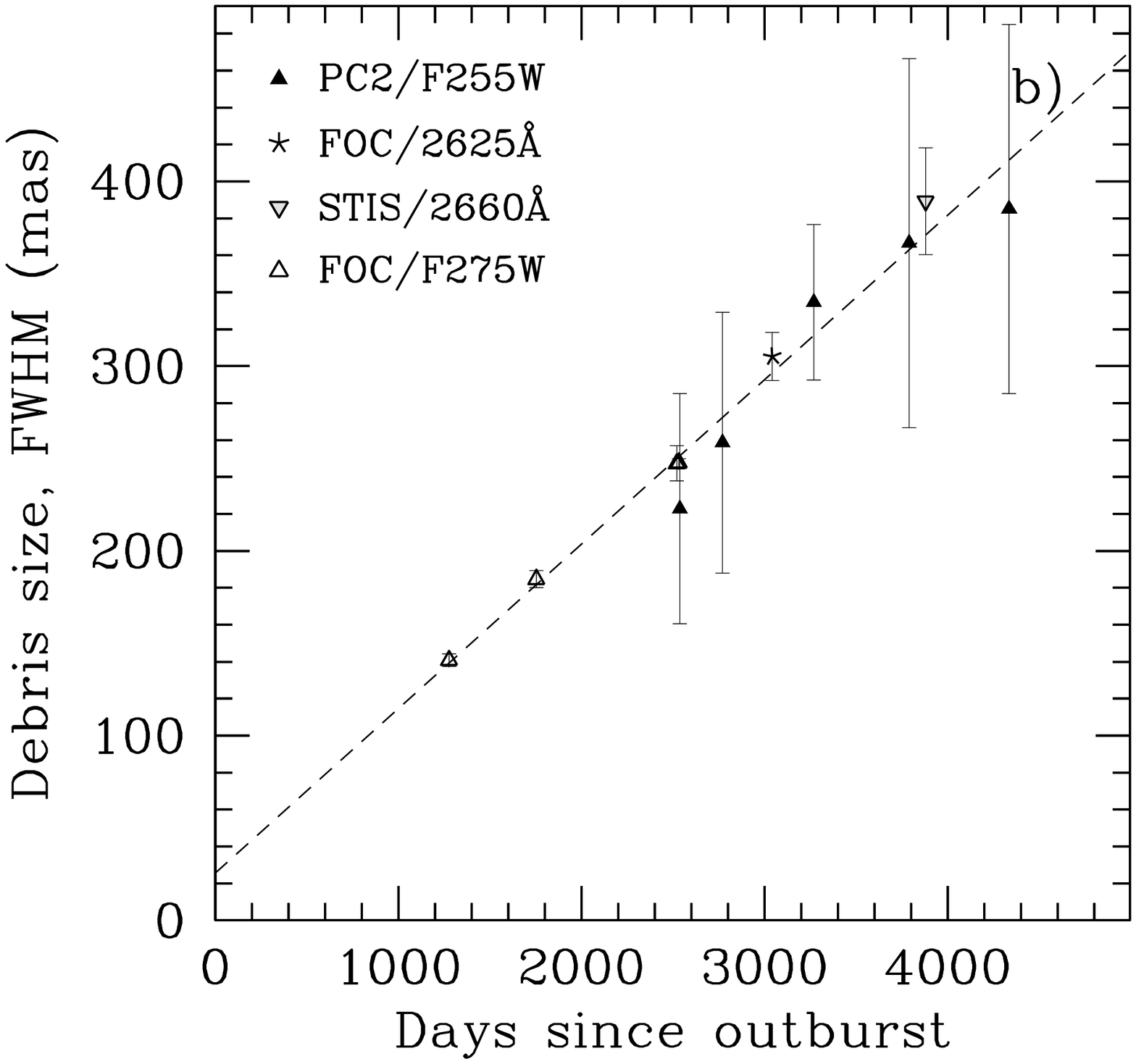}
}
\caption[]{
The apparent expansion of the ejecta of SN~1987A, as characterized by
the increase of the elliptical FWHM diameter (see text) of the envelope
with time.  (\emph{a}) The F501N (FOC) and F502N (PC2) filters.  Also
shown are measurements for the F547M and F555W (PC2) passbands and
measurements for the spectroscopic equivalent in the day~3034 FOC data
(central wavelength $\lambda_c$=4850\AA).  Error bars denote $\pm
1\sigma$.  The dashed line indicates the best-fit linear expansion rate. 
(\emph{b}) as (\emph{a}) for the F275W (FOC) and F255W (PC2) filters. 
Also shown are a FOC ($\lambda_c$=2625\AA) and a STIS
($\lambda_c$=2660\AA) data point for the spectroscopic equivalent of
these filters.  A linear fit (dashed) is overlayed. }
\end{figure*}
%

The best-fit linear expansion rates are (10.34\tpm0.84) mas~yr$^{-1}$
for the combined visible, and (16.27\tpm0.83) mas~yr$^{-1}$ for the
combined near-UV data.  For a distance to SN~1987A of 51 kpc (Panagia
\etal 1991) this corresponds to effective expansion velocities of
\tsim2500 and \tsim3930 km~s$^{-1}$, respectively.  These values match
the earlier results obtained by Jakobsen \etal (1994) from the FOC data
of days 1278, 1754 and 2522 to within the respective errors.

\subsubsection{Deviations from linear expansion?}

The FOC objective prism observations presented above leave little doubt
that the larger apparent size of the SN~1987A envelope in the near-UV is
due to the emission at these wavelengths being subjected to multiple
scattering and reprocessing in the optically thick resonance transitions
of \ion{Mg}{i}/\ion{Mg}{ii}/\ion{Fe}{ii} ions located in the outer,
faster moving regions of the ejecta as suggested by Wang \etal (1996)
and Chugai \etal (1997). 

However, as pointed out by the latter authors, in this case one does not
expect the outermost regions of the ejecta to grow in a strictly linear
fashion.  Specifically, in the case of conservative scattering in a
linearly expanding envelope displaying a power-law density profile
$\rho(r) \propto r^{-k}$, the radius at which optical depth unity is
reached is expected to increase with time as $r(t) \propto
t^{\frac{k-3}{k-1}}$. 

As discussed by Chugai \etal (1997), optical depth unity in the
\ion{Mg}{ii} line is reached in the very outmost regions of the ejecta
corresponding to expansion velocities of order $\sim9000$~km~s$^{-1}$,
where most models assume a very steep density gradient corresponding to
$k \simeq 9$.  Hence Chugai \etal predicted that outer boundary of the
envelope of SN 1987A should expand more slowly than linearly; \ie, as
$r(t) \propto t^{0.75}$ in the near-UV. 

As is evident from Fig.~5, the growth of the size of the SN1987A debris
-- when defined in terms of the FWHM of the images -- remains consistent
with linear expansion out to the last available data point (day 4336;
1999 Jan 7) in both the visible and the near-UV.  However, these
measures probe velocities in the range \tsim2500--4000 km~s$^{-1}$.  To
see the effect predicted by Chugai \etal we need to go 2--3 times
further out in the images, corresponding to angular radii of order
$\sim300$~mas and brightness levels $\sim10$\% of peak and fainter. 

For $k \simeq 9$ the anticipated deviation from linearity at these large
radii is only of order $\sim 11\%$ between day $\sim2500$ (and the last
high-quality FOC F275W imaging data) and day $\sim4000$ (and the last
PC2 F255W images).  Unfortunately, the available late-time \emph{HST}
data on SN~1987A do not allow us to detect such a relatively subtle
effect.  As is clear from Fig.~2, the outer regions of interest
beyond $\sim200$~mas radius in the FOC prism data are seriously hampered
by contamination from the dispersed images of star 3 and the
circumstellar ring.  Likewise, as is evident from Fig.~5(b) the S/N
ratio of the later day PC2 and STIS data are too low to permit a measure
of the image sizes at, say, 10\% peak at the required accuracy. 

We conclude that there is unfortunately no chance of detecting the
effect predicted by Chugai \etal in the late-time {\em HST} data --
and that the apparent linear expansion displayed in Fig.~5 therefore
does not challenge the resonance scattering explanation for the
wavelength dependence of the size of the SN~1987A debris.


\section{Conclusions}

We have presented \emph{HST} FOC near-UV objective prism observations of
the ejecta of SN~1987A taken on 1995 June 24 (3043 days after the SN
outburst).  We have combined these data with available archival FOC,
WFPC2, and STIS data to study the late-time expansion of the SN debris
over a period of \tsim8 years.  Our main findings are:

1) Provided the pronounced ellipticity of the SN image seen in the
visible data is taken into account, the available data are consistent
with the SN envelope having expanded linearly in time at all wavelengths
out to the last data point sampled (1999 January 7; day 4336). 

2) Throughout this expansion, the apparent size of the ejecta, expressed
in terms of the FWHM of the radial intensity profile, remained some
\tsim50\% larger in the near-UV than in the visible. 

3) The FOC near-UV prism spectrum reveals that the large spatial extent
of the SN~1987A image is confined to the 2350-2900~\AA\ wavelength
region containing the resonance lines of \ion{Mg}{i}, \ion{Mg}{ii} and
\ion{Fe}{ii}, thereby confirming the suggestion of Wang \etal (1996) and
Chugai \etal (1997) that the larger apparent size of the SN~1987A
envelope in the near-UV is due to multiple scattering and reprocessing
in these transitions in the outer regions of the envelope.

%
\end{document}